\documentclass{article}
\usepackage{graphicx}
\begin{document}
\title{Orientational Contribution to the Giant Electrostriction
Effect and Dielectric Permittivity in Relaxors}
\author{S. A. Prosandeev\\
(Physics Department, Rostov State University, Rostov on Don,
Russia)
 }
\maketitle
\begin{abstract}
A microscopic model is developed considering randomly oriented
polar regions stuck to the directions of random fields or
stresses. The application of an external field (stress) in this
model results in a dielectric and acoustic reply connected with a
polarization (strain) rotation.  Polarization (strain) fluctuation
contributions are also studied.
\end{abstract}
\maketitle
\section{Introduction}
Relaxors like PMN posses  very high values of the electrostriction
constant  and dielectric permittivity; the addition of PbTiO$_3$~
(PT) results in a large piezoelectric coefficient, which is
important for applications \cite{Park}. In the last decade these
properties have attracted the attention of theorists because of
fundamental reasons: first-principle computations revealed the
fact that, under the action of field, the polarization in the
steady state can rotate and the system can go through a series of
metastable states that results in a dramatic change of strains
\cite{Vanderbilt,Cohen}.

The polarization rotation mechanism seems to be general for
systems close to frustrations and, in relaxors, we have additional
facts supporting this mechanism. In particular this is the finding
of the monoclinic phase in PZT and PMN\cite{Noheda,Ye}. This phase
can be obtained by a continuous rotation of the polarization
vector between the tetragonal and rombohedral phases.

It is remarkable that the description of this finding within a
3-order-parameter scheme requires using a Landau expansion up to
the 12th-order \cite{Vanderbilt_Landau,Gufan} although from the
point of view of the polarization rotation mechanism this phase is
natural when going through a continuous path of the rotated
polarization vector.

The polarization rotation in relaxors is facilitated by the fact
that the local polarization can rotate only in comparatively small
volumes, which were called polar regions \cite{Smolenskii}. The
existence of such regions was recently convincingly established by
neutron scattering experiments, which showed the so called
"waterfall" phenomenon \cite{Vakhrushev,Gehring}, the essence of
which is the limited size of the polar regions leading to the
existence of the neutron scattering at large wave vectors only.

At high temperatures the polar regions are absent and the neutron
scattering shows the normal behavior; the dielectric permittivity
obeys the Curie-Weiss law. Obviously at these temperatures the
radius of the ferroelectric polarization fluctuations is small
compared to the heterogeneity radius but it increases with the
temperature decrease and at some temperature, known as the Burns
temperature, it reaches the heterogeneity radius.

At lower temperatures the correlation radius is unable to increase
further, because of the limitations imposed on the size of the
ferroelectric fluctuation in a disordered media, and a deviation
from the Curie-Weiss law is evidenced. At the same time the
neutron scattering at small wave vectors becomes flat and,
correspondingly, the optic modes can not be separated from the
acoustic ones; the central peak, which is usually attributed to
the elastic scattering, is growing. These features exactly
correspond to the idea of the polarization rotation but in this
case local polarization fluctuations of the rotational kind (they
can be also called the hydrodynamic polarization fluctuations in
order to distinguish them from the ordinary ferroelectric
fluctuations) in the finite-size polar regions become important as
the correlation radius of the ordinary ferroelectric fluctuations
had been already saturated. Just these fluctuations, in our
opinion, are responsible for the central peak, waterfall
phenomenon dielectric losses and, as we will see, also for the
strong electrostriction and piezo- effects.

If one imagined for a moment that the polarization vector
direction is fully frustrated in the space then such a system
would be unstable due to the degeneracy of the energies of the
states with different polarization vector directions; the
transverse dielectric susceptibility of such a system would
diverge. However this system can be stabilized by fields or
stresses.

In relaxors there are special conditions for such fields
(stresses) but they are local and random \cite{Viehland}. The
origin of these fields (stresses) is in the random distribution of
the ions with different charges (sizes) over the similar sites.
The stabilization of the nanoscale polar regions by the random
fields and stresses proves to be one of the most important
features of the relaxors, in our opinion.

In fact one can consider the polarization vector directions
frustrated but stuck to the local random fields, which can have
different magnitudes in different crystallografic directions. It
implies that, from the first glance, the polarization vector
conserves its magnitude when turning aside but the energy of this
dipole is different in different directions due to the interaction
of the polarization with the internal local fields (stresses).

Thus the idea of the polarization rotation \cite{Cohen}
supplemented by the idea of the random fields (stresses)
\cite{Viehland} seems to be rather promising and further studies
in this direction are necessary. As this idea is general for all
systems close to frustrations the main features and consequences
stemming from this mechanism should be described within a clear
and simple (possibly analytical) model in order to use it for
qualitative explanations of new experimental facts and
predictions.

The present study is kind of an endeavor to develop such a model
within a random field approach and based on the results of
first-principle computations \cite{Vanderbilt,Cohen}. Some
preliminary results applied to KTa$_{1-x}$Nb$_x$O$_3$~ as an
example were published in Ref. \cite{random}. At the first stage
the polarization rotation contribution to the dielectric reply
will be considered. Then this approach will be extended to the
case of the acoustic reply.

\section{Orientable Polar Regions} \label{Orientable}

Consider random fields \textbf{e} and merged to their directions
local dipole moments ${\rm {\bf \mu }}$. In the field ${\rm {\bf
E}} = {\rm {\bf E}}_{\rm {\bf 0}} + \eta {\rm {\bf P}}$, where
\textbf{E}$_{0}$ is the external field and \textbf{P} being the
polarization, the dipole moments are directed along \textbf{E} +
\textbf{e}.

We assume further that the local polarization has been already
appeared in the polar regions and, due to this fact, the action of
the field results in the redirection of the dipole moments without
changing their magnitudes. It is true if $k_BT\ll\mu e$. Thus this
is a purely geometrical problem: one should find the new
directions of the local fields at each of the points in space and
redirect the local dipole moments from the previous directions
determined by only the random fields to the new directions, which
are the vector sums of the random and external fields. The final
result should be averaged over the points in the space.

It is obvious that in the presence of the external field the
average polarization is not longer zero. It can be found from the
following analytical expression

\begin{equation}
\label{polariz} \left<P\right>=\frac{n\mu}{2}\int\limits_0^\pi
p(\theta,E) \sin \theta d \theta
\end{equation}
\noindent where
\begin{equation}
 p(\theta,E)=\frac{ E + e\cos \theta }{ \sqrt {E^2 +
 e^2 + 2Ee\cos \theta }}
\end{equation}
The integration gives
\begin{equation}
\left< P \right> = \left\{
  \begin{array}{*{20}c}
          {n\mu (1 - e^2 / 3E^2)} \hfill & {E > e} \hfill \\
          {2n\mu E / 3e} \hfill          & {E < e} \hfill \\
  \end{array}
  \right.
\end{equation}

It follows from this result that the susceptibility of
nonineracting polar regions can be found from the expression:

\begin{equation}
\label{suscept} \chi _0 = \left. {\frac{1}{\varepsilon _0
}\frac{dP}{dE}} \right|_{E = 0} = \left\{ {{\begin{array}{*{20}c}
 {2n\mu / 3e\varepsilon _0 } \hfill & {E < e} \hfill \\
 {2n\mu e^2 / 3\varepsilon _0 E^3} \hfill & {E > e} \hfill \\
\end{array} }} \right.
\end{equation}

\noindent where $n$ is the dipole (heterogeneity) concentration.
The temperature dependence of the dipole moment is determined by:
$\mu = \mu _0 \tanh \left[ {\left( {{\rm {\bf \mu }}_0 {\rm {\bf
e}}} \right) / k_B T} \right]$ where ${\rm {\bf \mu }}_0 $ is the
dipole moment magnitude. The Free energy of the noninteracting
dipoles can be found by the integration of the dielectric
susceptibility:

\begin{eqnarray}
F=\left\{
\begin{array}{*{20}c}
{\frac{3e}{4n\mu}\left(P-P_0\right)^2-EP} & {E<e} \\
{\frac{8n \mu e^2}{3}{\sqrt{ P_1 - P}}-EP} &{ E>e} \\
\end{array}\right.
\end{eqnarray}

\noindent where $P_0$~ and $P_1$~ are the lowest and highest
values of the polarization in the field correspondingly. The $P$~
value is the polarization magnitude as the direction of the field
coincides with the direction of the polarization.

The Hamiltonian, which takes into account the local field effects
and interactions among the polar regions, can be written in the
form

\begin{eqnarray}
 \label{eq6}
 H = \frac{1}{2}\left(
           {\frac{3e}{2n\mu}
           - \eta }
     \right)\delta P^2 +
                               \nonumber \\
 + \frac{1}{4}\tilde {\beta
 }P^4 + \frac{1}{6}\xi P^6 + \upsilon \left( {\nabla P} \right)^2 -
 EP
\end{eqnarray}

\noindent where $\tilde {\beta } = \beta - 4q / c$, $q$~ is the
electrostriction constant and $c$~ being the elastic constant. A
large value of the electrostriction constant leads to negative
values of the nonlinearity constant $\tilde {\beta }$ and,
consequently, to the phase transition of the first order. Hence
the orientable polarization in relaxors results in an additional
contribution to the dielectric permittivity, which can be rather
large at small random field values.

To take into account the scattering of the random field magnitude
we used the following distribution function for a reorientable
part of the random fields \cite{Glinchuk}

\begin{equation}
\label{distr} f(e) = \frac{1}{\left( {\sqrt \pi a} \right)^3}e^{ -
\left| {{\rm {\bf e}} - \eta {\rm {\bf P}}} \right|^2 / a^2}
\end{equation}

By integrating the expression for the dielectric susceptibility
(\ref{suscept}) with this function we obtained at $E < e$

\begin{eqnarray}
\label{eq8} \chi _0 = \frac{4n\mu }{3\varepsilon _0 \eta
P}erf\left( {\eta P / a} \right) \approx \nonumber \\
\approx \frac{4n\mu }{3\sqrt \pi
\varepsilon _0 a}\left[ {1 - \frac{\eta ^2P^2}{3a^3} + ...}
\right]
\end{eqnarray}

\noindent It is seen that the bare susceptibility decreases with
the width of the distribution function (\ref{distr}) and with
polarization $P$.

We regard the frequency dependence of the dielectric permittivity
to potential barriers separating different positions of the random
fields. One can consider the random fields coupled to the soft
vibrations and these random fields can be polarized by external
field according to the distribution function (\ref{distr}).

Consider the case when the polar regions are embedded in an $ac$~
field. Polarization can be found as
\begin{equation}
P_z= n \mu^2 F(T) \tilde{E}_z
\end{equation}
\noindent where $F(\omega ) = \left[4k_BT\left(1 - i\left(\omega
\tau\right)^{1 - \alpha }\right)\right]^{-1}$; $\tilde{E}_z = E_z
+ \lambda P_{TO}$~ is the local field, $E_z$~ is an external field
and $P_{TO}$~ is the polarization connected with the soft mode and
$\lambda$~ is a coupling constant. The corresponding contribution
to the dielectric susceptibility can be written as

\begin{equation}\label{chiLi0}
\chi=\frac{1}{\varepsilon_0}\frac{dP_z}{dE_z}=\frac{n
\mu^2}{\varepsilon_0} F(T) \left(1+\lambda \varepsilon_0
\chi_{TO}\right)
\end{equation}

\noindent Now one can obtain

\begin{equation}\label{coupling}
\chi_{TO}=\frac{1}{\varepsilon_0} \frac{1+\lambda n \mu^2
F(T)}{A(T)-\lambda^2n \mu^2 F(T) }
\end{equation}

\begin{equation}
A(T)=\alpha +3 \beta P_{TO}^2 + 5 \gamma P_{TO}^4 - q e
\end{equation}

\noindent Here $q$~ is an electrostriction constant and $e$~ is
strain. It is seen that the dielectric susceptibility connected
with ferroelectric fluctuations and polar regions is enhanced due
to their mutual coupling; the critical temperature is shifted to
higher temperatures; the frequency dispersion is also enhanced at
the phase transition point.

\section{The dependence of strain on the field}

The strained samples become softer in the direction of the
elongated axis due to the decrease of the short-range forces. This
effect leads to the ordinary electrostriction coupling $-qP^2e$.
The electrostriction coupling can trigger even a phase transition
if the stress is large enough. Besides this, ordinary, effect,
which has been already well known, we consider here another
effect, the alignment of the polar regions due to the stress. The
polar regions can be aligned due to the enhancement of the
indirect dipole-dipole interaction over the soft mode and because
of the enlargement of the probability to occupy the wells at the
elongated axis. At first we consider the former possibility and
then the latter.

To take a look at the former effect one can use expression
(\ref{coupling}), from which it is seen that the Debye reply
$F(T)$~ in the denominator is enhanced by the factor
$\lambda^2/A(T)$. Hence the decrease of $A(T)$~ due to
electrostriction coupling increases the polar region contribution
to the dielectric susceptibility and can result in the alignment
of the dipoles.

Further we analyze the latter mechanism in the same manner as in
the previous section. One can imagine that each polar region
produces a stress along the axis of the local polarization and
this stress is proportional to the square of the local dipole
moment. In the paraelectric phase, on average, the sample is cubic
but, nevertheless, the local stress does exist due to the
existence of the local dipole moments. When the sample is poled
the average stress is not longer zero because of the "alignment"
of the strained polar regions in the poling field (this mechanism
could be called the "strain rotation" mechanism, which resembles
the "polarization rotation mechanism").

The average strain in the field {\bf E} can be found from

\begin{eqnarray}
s \sim \left<P^2\right>=n \mu^2
\int\limits_{-\pi}^{\pi}{d^2(\theta,E) \sin\theta d\theta}=
\nonumber \\ = \frac{n\mu^2}{16eE^3}[12eE^3-4e^3E+(E^2-e^2)^2\ln
\frac{(e+E)^2}{(e-E)^2}]
\end{eqnarray}

At small fields this average behaves quadratically with the field

\begin{equation}\label{square}
\left<P^2\right>=\frac{n\mu^2}{3}\left(1+\frac{E^2}{10e^2}+
...\right)
\end{equation}

In the case of the poled sample one should replace $E$~ in Exp.
(\ref{square}) with $E+E_p$~ where $E_p$~ being the internal field
arising due to poling the sample. At small poling fields it
results in

\begin{equation}
\left<P^2\right>=\frac{n\mu^2}{3}\left(1+\frac{E_pE}{5e^2}+ ...\right)
\end{equation}

Hence, in ferroelectrics one has

\begin{equation}
s \sim \frac{n\mu^2}{3}\frac{E_pE}{5e^2}+ ...
\end{equation}

It is seen that this contribution to the strain is especially
large in the case of small magnitudes of the random fields (which,
nevertheless, should satisfy the inequality: $\mu e \gg k_B T$).

The piezoeffect coefficient, $ds/dE$, increases with the poling
field $E_p$~ but at large poling fields the stain saturates. If
one takes into account the interaction among the dipoles then an
abrupt change of the polarization at some field is possible and
this change can cause an abrupt change of the stress.

Thus the orientational polarization and strain can contribute much
to the average macroscopic dielectric susceptibility,
electrostriction constants and piezoeffect in the case if  the
sample contains orientable polar regions stuck to small random
fields (stresses).

Notice that the question about the conditions for the appearance
of the polar regions is a hard question because of the
depolarization field. However in the presence of movable charge
carriers or due to a cooperative appearance of self-compensated
polar regions these conditions become easier.

\section{Polarization fluctuations}

Here we study longitudinal polarization \textit{fluctuations}
caused by the polarization rotations. We will see that the
polarization fluctuations in the polar regions can provide a large
additional contribution to the dielectric permittivity. We
consider the polarization field \textbf{P}(\textbf{r}) directed
along the local field \textbf{E}$_{l}$. In a weak transverse field
$\delta E_{l \bot } $ the transverse polarization has the form:
$\delta P_{l \bot } = P_1 \delta E_{l \bot } / E_l $ where $P_{1}$
is the polarization magnitude inside the polar region. Hence the
transverse susceptibility is $\chi _{l \bot } = P_1 / \varepsilon
_0 E_l $.

The transverse polarization appeared at one of the space points
causes the appearance of a transverse polarization at nearest
space points. In order to describe the profile of the polarization
field fluctuation one can write the Hamiltonian with taking into
account the gradient term

\begin{equation}
\label{eq1} F_{l \bot } = \int {\left( {\chi _{l \bot }^{ - 1}
P_{l \bot }^2 + c\left( {\nabla P_{l \bot } } \right)^2 - E_{l
\bot } P_{l \bot } } \right)dV} .
\end{equation}

The corresponding correlation function is the Ornstein-Cernike
function

\begin{equation}
\label{eq2}
 < \delta P_{l \bot } (0)\delta P_{l \bot } ({\rm {\bf r}}) > = \frac{k_B
T}{4\pi cr}\exp \left( { - \kappa r} \right)
\end{equation}

\noindent where $\kappa ^2 = \left( {c\chi _{l \bot } } \right)^{
- 1} = \varepsilon _0 E_l / cP_1 $. i In order to find the
longitudinal susceptibility we use the mathematical trick
suggested in (\cite{Patashinski}): we employ the condition of the
polarization magnitude conservation $\delta P_{l\parallel } =
\left( {\delta P_{l \bot } } \right)^2 / 2P_1 $ and find that

\begin{eqnarray}
\label{eq3} \chi _{l\parallel } = \frac{d}{dE_l }P_{l\parallel } =
\frac{1}{2P_1 }\frac{d}{dE_l }P_{l \bot }^2 = \nonumber \\
\frac{1}{2P_1 }\left. {\frac{d}{dE_l }\left\langle {P_{l \bot }
\left( 0 \right)P_{l \bot } \left( {\rm {\bf r}} \right)}
\right\rangle } \right|_{{\rm {\bf r}} = 0}
\end{eqnarray}

This leads to

\begin{equation}
\label{eq4} \delta P_{l\parallel } = \frac{k_B T}{8\pi \left(
{cP_1 } \right)^{3 / 2}\left( {\varepsilon _0 E_l } \right)^{1 /
2}}\delta E_l = \frac{a}{\sqrt {E_l } }\delta E_l
\end{equation}

Now, by averaging the result over the random field directions and
projecting the result onto the z axis we have

\begin{eqnarray}
\label{eq5}
  \chi_{\vert \vert } = \frac{2a}{7eE^3}
  \{
      \left| {E+e} \right|^{7 / 2}
    - \left| {E-e} \right|^{7 / 2}
                                            - \nonumber \\
              - \frac{7}{6}
      \left( {E^2 + 2e^2} \right)
          \left[ \left| {E + e}\right|^{3 / 2}
                   - \left| {E - e}\right|^{3 / 2}
          \right]
  \}
\end{eqnarray}

\noindent where $\chi _{\vert \vert } = dP / dE$~ and $P$~ is the
projection of the polarization on the direction of the external
field. The longitudinal contribution has a finite value at zero
external field ($4a / 3\sqrt e $, which increases with the random
field magnitude decrease), increases in small fields
quadratically, then reaches a maximum and then decreases as $E^{ -
1 / 2}$. Thus the polarization rotation mechanism results in a
large fluctuation contribution to the dielectric permittivity.

In similar way one can consider an acoustic reply. In this case
the strain fluctuation should be considered instead of the
transverse polarization fluctuation. The physics of these
fluctuations is similar as the stress like the field can provide a
polar region reorientation. This reorientation appears because in
a tetragonally distorted crystal the probability of the occupation
of the wells lying at the elongated axis differs from the
probability to occupy the wells at the other axes.

\section{Easy domain wall movement}

The polarization rotation mechanism can result in facilitating the
domain wall movements \cite{Bruce} in multidomain ferroelectric
samples. Indeed consider domain walls at random directions
stabilized by the electrostatic potential produced by the
surrounding domains. Let {\bf n} be a unit vector perpendicular to
the wall. If one shifts the wall along {\bf n} by a small distance
then a force appears returning the wall to the previous place:
$F=-kx$~ where $k$~ is a spring constant. Consider $\theta$~ be an
angle between the external field $E$~ and {\bf n}. In this case
the average polarization in the field $E$~ is:

\begin{equation}
P=\frac{Z^{*2}E}{k}\int\limits_0^{\pi}{ \cos^2\theta \sin \theta d
\theta }= \frac{Z^{*2}}{3k}E
\end{equation}

\noindent where $Z^*=PS$~, $P$~ is polarization magnitude and $S$~
being the domain width. It is seen that the dielectric
susceptibility here is determined mainly by the small spring
constant value, which is governed by small potential barriers
between the dipoles oriented in different possible directions. By
employing the same assumptions as above regarding the average
strain one can obtain

\begin{equation}
s \sim \frac{Z^{*2}E^2}{2k^2}\int\limits_0^\pi{\cos^4\theta \sin
\theta d \theta}=\frac{Z^{*2}E^2}{5k^2}
\end{equation}

\noindent which for poled samples results in

\begin{equation}
s \sim \frac{Z^{*2}E_pE}{5k^2}
\end{equation}

Again one can see that a large strain can appear because of the
fact that the system is nearly frustrated at the equilibrium
position and it is stabilized by only weak random fields and
stresses. Such a a description is rather similar to the
orientational mechanism, which we considered above. In both cases
we have the result, which increases with the random field (stress)
constants decrease. However the fact that these constants have
finite values is significant as just these fields (stresses)
stabilize the considered state of the system where we have the
system at the frustration point.

This consideration can hold only at comparatively small deviations
from the equilibrium state. In the case if the fields are large
then the domains can change their stable configuration abruptly,
which can be considered as the relaxation of the random fields
(stresses) to new positions with a new distribution function (see
Section \ref{Orientable}). Coupling between this relaxation and
polarization fluctuations can produce a cooperative dynamics.

\section{Discussion}

The present study has shown that the existence of the nanoscale
orientable polar regions in relaxors can provide an additional to
the ordinary ferroelectric fluctuations contribution to the
dielectric permittivity, electrostriction and piezo effects.
Besides this a fluctuation contribution was shown to be also
large. These data shed light to the experimental fact found that
the relaxors have extremely large electrostriction and piezo-
coefficients.

A support of these data was given also by the study of solid
solutions of quantum paraelectrics like KTN \cite{Toulouse}, which
showed a softening of the elastic constants of KTN in the {\it
paraelectric phase} well above of the forroelectric phase
transition due to the impurity cluster formation. A rotational
mechanism employed in Ref. \cite{Pattnaik} gave a good explanation
of the nonlinear electrostriction behaviour in this solid
solution.

Thus the nanoscale formations in the disordered materials can play
an important role in determining the physical properties of these
materials that can be used in practice.

Notice that the Ising model misses these effects as it considers
usually two states while in the explanation shown above the
3-dimensional structure of the polar regions is significant as
just this structure allows the orientational polarization and
fluctuations. The superparaelectric model is also very different
as it considers the thermally fluctuating dipoles with the average
dipole magnitude dependent on temperature and this is the main
point in this model. We showed above that such a model is unstable
with respect to the transverse polarization fluctuation and one
\emph{must} to consider fields or stresses stabilizing the polar
regions and their fluctuations.

The present model considers the polar region dipoles stuck to the
random directions of internal electric fields (stresses), which is
significant for the final result. We have shown that being close
to the instability at small random field values one can have a
rather large contribution to the dielectric permittivity,
electrostriction and piezo- effects due to the polarization and
stress rotations.

The rotational mechanism is not the only possible mechanism for
the large polarization fluctuations in relaxors. Another
possibility is the dynamics of the domain walls separating the
nanoscale domains \cite{Bruce}. This mechanism is similar to the
one discussed in the present paper in the point that we  have a
system close to frustration and actions of small fields or
stresses shift the equilibrium coordinates sufficiently. In the
orientational mechanism these coordinates are the angles but in
the domain walls case these are the positions of the walls.

In any case we have to have random fields or stresses in order to
stabilize such a system. If these fields are small then the
dielectric and acoustic replies will be large. The presence of
nonlinearities can provoke a phase transition in the field or
under the stress.

\section{Acknowledgement}

The author appreciates discussions with A. Bokov. The study was
supported by RFBP, grant \# 01-02-12069.

\end{document}